# Generalized Hyper-Ramsey spectroscopy in an optically dense medium

K.A. Barantsev[1], T. Zanon-Willette[2], A.N. Litvinov[1]

[1]Peter the Great St.Petersburg Polytechnic University, 195251, Russia, Saint-Petersburg

[2] Sorbonne Université, Observatoire de Paris, Université PSL, CNRS, LERMA, F-75005, Paris, France

e-mail: kostmann@yandex.ru, thomas.zanon@sorbonne-universite.fr, andrey.litvinov@mail.ru

## ABSTRACT

In this work, the peculiarities of Ramsey resonance and its sensitivity to the light shift from an optically dense medium of cold atoms are investigated. We considered different composite pulse protocols for clock spectroscopy such as hyper-Ramsey, modified and generalized hyper-Ramsey schemes. Shapes of resonances and error signals changes significantly due to the processes of absorption and dispersion in the atomic medium. The dependence of the position of the central fringe resonance with a residual uncompensated light shift of the atomic transition is theoretically studied when taking into account the attenuation of the radiation intensity in the medium. The change in pulses area of the hyper-Ramsey protocol allows us to suppress the sensitivity of the clock resonance position to the residual light shift for a certain length of the medium. It is shown that using a combination of generalized hyper-Ramsey error signals allows us to suppress the sensitivity to the light shift for any length of the medium. Also we analyzed the effect of spontaneous decay of high atomic state on the light shift sensitivity of the composite pulses schemes.

## 1. Introduction

Studies on magnetic resonance were started about 80 years ago [1]. In this work, I.I. Rabi, by improving technology, managed to increase the resolution and obtain a lot of new information not only about the atomic and molecular structure, but also about the atomic properties [2]. Since a two-level atom is similar to a particle with a half-full spin in a magnetic field, the basic dynamic equations describing the evolution of a variable two-level atom almost coincide with the equations describing spins. Consequently, the Bloch formalism for the spin vector, developed to describe magnetic resonance, can be applied to optical resonance problems [3].

Optical resonance between two (primary and excited) quantum levels can be used as a frequency reference for a new generation of ultra-stable and accurate atomic clocks. Optical frequency standard can be realized with single ions [4-6], neutral atoms in optical lattices [7-9], and more recently with the ultraviolet transition in the nuclear thorium-229 core [10-12]. Using the optical transition as a reference allowed us to achieve atomic clock accuracy at the relative level of $10^{-18}$. Such optical atomic clocks open new opportunities for measuring the drift of fundamental constants, checking the laws of quantum electrodynamics [13], cosmological gravimetry [14], and detecting dark matter [15]. It is expected that such an active research in the field of time and frequency metrology will open soon an access to a relative accuracy below the $10^{-18}$ [16, 17].

In 1949, N.F. Ramsey proposed using excitation of atom by a sequence of two radiation pulses (instead of a single one) separated by a free evolution time [18]. Such a scheme makes it possible to reduce any perturbation induced by the probing electromagnetic field compared to the Rabi interrogation. This method quickly found application in the field of quantum frequency standards, first of all, these are microwave frequency standards [19,20]. Later this method was used in optical frequency standards and atomic interferometry by using a sequence of three, four (and more) interaction pulses [21,22].

When an atom interacts with laser radiation in the standard Ramsey scheme, due to the presence of non-resonant atomic transitions, a light shift occurs (the Stark effect), which depends linearly on the intensity. In high-precision optical clocks, it is necessary to know the exact position of the resonance, but the intensity of the interrogation laser fluctuates. Random changes in intensity limit the precise determination of the position of the resonance line. It is leading to limit of the accuracy of optical clocks. In 2010, and 60 years after the original technique of separated oscillating fields, the hyper-Ramsey survey method was proposed [23]. The essence of the scheme is to use a sequence of time-separated interactions with the probe laser acting as composite pulses, which can have different durations, frequencies and phases. For certain parameters of the pulse train, the dependence of the position of the resonance line as a function of the light shift of the resonant transition is similar to the shape of a cubic parabola.
From this, it follows that there exists a region near the resonance, where the position of the resonance depends very weakly on the light-shift correction. Thus, the use of such an interrogation method allows to increase accuracy of optical clocks by removing the need to extrapolate any light-shift correction at zero power with precision. Experimental demonstration of the hyper-Ramsey method [24] has given a new impetus to research in this area. Thus in [25-27] the influence of various parameters of pulses, frequencies and laser phase-steps on the resonance position was studied. In [28], the effect of fluctuations of the probe laser field for the hyper-Ramsey spectroscopic scheme was studied based on a simple noise model.

When using a cold atomic ensemble as the basis for an optical clock, it becomes obvious that the concentration of atoms can manifest itself by taking into account collective effects. In this case, the medium becomes optically dense. In such a medium, the shape of the resonance line changes in comparison with an optical thin medium [30,31].

In the above papers [23–28], including the review [29], an optically thin medium was considered to study various interrogation protocols and their influence on the clock frequency-shift. In [32], it was shown that the presence of an optically dense medium distorts the shape of the CPT-Ramsey resonance line. It would be logical to expect that when a hyper-Ramsey interrogation scheme is implemented in an optically dense medium, new features will appear, in particular, the shape of the curve describing the dependence of the light shift on the detuning of the laser field will change. This paper is devoted to the study of the hyper-Ramsey scheme for a "two-level" atom in an optically dense medium. The paper studies the fundamental physical effects and discusses their practical field of application.

## 2. Mathematical model and basic assumptions

Let's consider interaction of the pulsed laser radiation with the atomic ensemble. The ensemble consists of identical motionless atoms, one of the transitions of which ($|1\rangle \leftrightarrow |2\rangle$ with the frequency $\omega_{at}$) is close to the carrier frequency $\nu$ of the laser field, so the detuning is $\delta = \nu - \omega_{at} \ll \omega_{at}$. The ensemble has the length L along z axis of the propagation of the laser radiation (Fig.1). The mean free path of a photon is comparable to the length of the ensemble, therefore the atomic medium can be optically dense ($n_a \sigma L \sim 1$, where $n_a$ is the atomic concentration, $\sigma$ is the scattering cross section of photons by atoms). The ensemble is assumed to be quite dilute, such that the wavelength of the incident radiation is less than average interatomic distance ($n_a \lambda^3 < 1$, where $\lambda$ is the wavelength of the incident radiation). It allows us to neglect the effects of recurrent light scattering [33-36] and to consider the interaction of each atom with radiation independently in terms of quantum correlations. However, the interaction of radiation with each atom of the ensemble is not completely independent due to its optical density. The radiation incident on the atoms of the far layers of the ensemble depends on the state of the atoms of the nearby layers, which is a manifestation of the collective scattering of light [37,38].

The electric field of the laser is

$$\mathbf{E}(z,t) = \mathbf{e}_p E_0(z,t) e^{i(kz-\nu t)} + c.c., \qquad (1)$$

where $\mathbf{e}_p$ is the unit vector of polarization, $E_0(z,t)$ is the complex amplitude, $\nu$ is the carrier frequency, $k$ is the wave number. Further it is assumed that the direction of polarization does not change due to the isotropy of the medium, or changes negligible with weak anisotropy due to magnetic field, therefore $\mathbf{e}_p$ does not depend on longitudinal coordinate z. The semiclassical approach can be applied for the considered intensities of the laser field. Within this approach, the radiation is described classically but the atoms and their interaction with field is quantum. The Hamiltonian is

$$\hat{H} = \hat{H}_0 + \hbar \hat{V}, \qquad (2)$$

where $\hat{H}_0 = \sum_n \varepsilon_n |n\rangle\langle n|$ is the atomic Hamiltonian in the absence of the laser field, $\varepsilon_n$ are the energies of the atomic levels, n=1,2. $\hat{V}$ is the interaction operator which in rotating wave approximation is

$$\hat{V} = -\frac{\Omega^*}{2} e^{i(\nu t - kz)} |1\rangle\langle 2| - \frac{\Omega}{2} e^{-i(\nu t - kz)} |2\rangle\langle 1|, \qquad (3)$$

where $\Omega = 2(\mathbf{d}_{21} \cdot \mathbf{e}_p) E_0 / \hbar$ is the Rabi frequency, $\mathbf{d}_{21}$ is the matrix element of the dipole momentum operator.

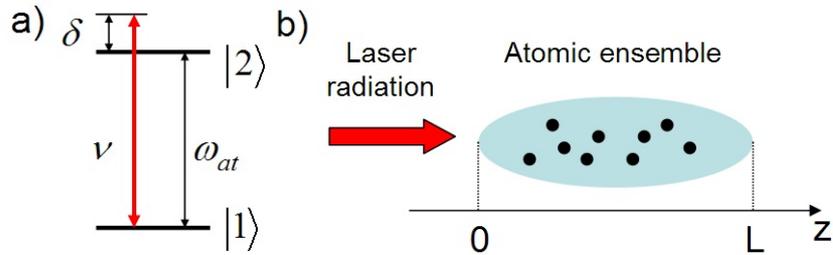

FIG.1. (Color online) (a) Interaction of the electromagnetic field with the frequency ν with the atomic transition with the frequency $\omega_{at}$. (b) Optically dense along z axis atomic ensemble.

Atomic system is described by Liouville's equation for the one-atom density matrix $\hat{\rho}$:

$$\frac{\partial \hat{\rho}}{\partial t} = -\frac{i}{\hbar}\left[\hat{H}, \hat{\rho}\right] + \hat{\bar{R}}\{\hat{\rho}\}, \qquad (4)$$

where $\hat{\bar{R}}\{\hat{\rho}\}$ is the super operator of spontaneous decay of the excited level. Propagation of the radiation in approximation of the plane wave front is described by one dimensional wave equation for the complex amplitude:

$$\frac{\partial E_0(z,t)}{\partial z} + \frac{1}{c}\frac{\partial E_0(z,t)}{\partial t} = 4\pi i P_0(z,t)k, \qquad (5)$$

where c is the light velocity, $P_0(z,t)$ is the slowly varying amplitude of the atomic polarization which is expressed through the atomic density matrix and the dipole moment operator $\hat{\mathbf{d}}$:

$$\mathbf{P}(z,t) = n_a \, \mathrm{Sp}\left(\hat{\rho}\hat{\mathbf{d}}\right). \qquad (6)$$

Let's substitute the expressions (2), (3) in the equation (4) and the expression (6) in the equation (5). After that, separate the rapidly oscillating factor in the off-diagonal elements of the density matrix $\rho_{12} = \tilde{\rho}_{12} e^{i(vt-kz)}$. Finally, we obtain the Maxwell-Bloch set of equations which describes the dynamics of the density matrix and propagation of the field:

$$\dot{\rho}_{11}(z,t) = -i\frac{\Omega(z,t)}{2}\tilde{\rho}_{12}(z,t) + i\frac{\Omega^*(z,t)}{2}\tilde{\rho}_{21}(z,t) + \gamma\rho_{22}(z,t), \qquad (7)$$

$$\dot{\rho}_{22}(z,t) = i\frac{\Omega(z,t)}{2}\tilde{\rho}_{12}(z,t) - i\frac{\Omega^*(z,t)}{2}\tilde{\rho}_{21}(z,t) - \gamma\rho_{22}(z,t),$$

$$\dot{\tilde{\rho}}_{12}(z,t) = \left(-i\delta(z,t) - \gamma/2\right)\tilde{\rho}_{12}(z,t) - i\frac{\Omega^*(z,t)}{2}(\rho_{11}(z,t) - \rho_{22}(z,t)),$$

$$\frac{\partial \Omega(z,t)}{\partial z} + \frac{\partial \Omega(z,t)}{c\partial t} = \frac{4\pi i n_a |d_{12}|^2 k}{\hbar}\tilde{\rho}_{21}(z,t).$$

Here the equation of field propagation is written for the Rabi frequency. γ is the spontaneous decay rate of the excited level, which is expressed through the matrix element of dipole moment as $\gamma = 4\omega_{at}^3 |d_{12}|^2 / 3\hbar c^3$ according to the theory of spontaneous emission of atoms.

In the set of equations (7) the laser detuning $\delta(z,t)$ depends on coordinate and time because, in general, the laser radiation interacts with other atomic transitions, which causes a light shift of a resonant transition $|1\rangle \leftrightarrow |2\rangle$. Suppose that under the action of a rectangular laser pulse with the amplitude $\Omega_0$, there occurs the shift $\Delta_{LS}$ in the frequency of the transition. This shift takes place only under the action of pulses, and during the free evolution time it is zero. In this way, if the laser has the detuning δ from the unperturbed transition when switching off the probe laser, the detuning will acquire an additive frequency-step during the pulse action as $\delta + \Delta_{LS}$. However, laser pulses, when passing through an optically dense medium, change their amplitude and cease to be rectangular. Since the light shift is proportional to the intensity of the radiation, the detuning of the laser field at the moment of time t at the point of space z takes the form

$$\delta(z,t) = \delta + \Delta_{LS} \frac{|\Omega(z,t)|^2}{|\Omega_0|^2}, \qquad (8)$$

where $\Omega_0$ is the amplitude of a rectangular pulse at the entrance to the medium.

## 3. Discussion of the results
### 3.1 Analyze of the hyper-Ramsey resonances

When using the classical Ramsey scheme of interrogation, a sequence of two identical pulses with the area $\pi/2$ acts on atoms (Fig.2(a)). Change in the detuning according to formula (8) for z=0 is shown in Fig.2(b).

In atomic clocks, the sensitivity of the position of the Ramsey resonance to fluctuations of the light shift $\Delta_{LS}$ of an atomic transition directly affects their stability. Let's consider the light shift $\Delta_{LS}$ as a free parameter which in a real experiment is determined by the degree of interaction of laser radiation with non resonant atomic transitions and the intensity of radiation. The position of the resonance will be understood as the position of its extremum S on the axis of frequency. For the classic Ramsey scheme of interrogation the dependence $S(\Delta_{LS})$ is linear in the vicinity of the point $\Delta_{LS}=0$. In the work [23], it was proposed to use a more complex sequence of pulses, which reduced sensitivity of the position of the resonance to the light shift (hyper-Ramsey method). The sequence of the pulses in the Fig.2(c) makes it possible to reduce the dependence $S(\Delta_{LS})$ near zero to cubic, that significantly reduces the sensitivity of the resonance to variations of the light shift. In this sequence, the first pulse has an area $\pi/2$, and the second composite pulse consists of two parts with the areas (-$\pi$) and $\pi/2$. Dependence of the detuning on time for such sequence of the pulses according to formula (8) is shown on the Fig.2(d).

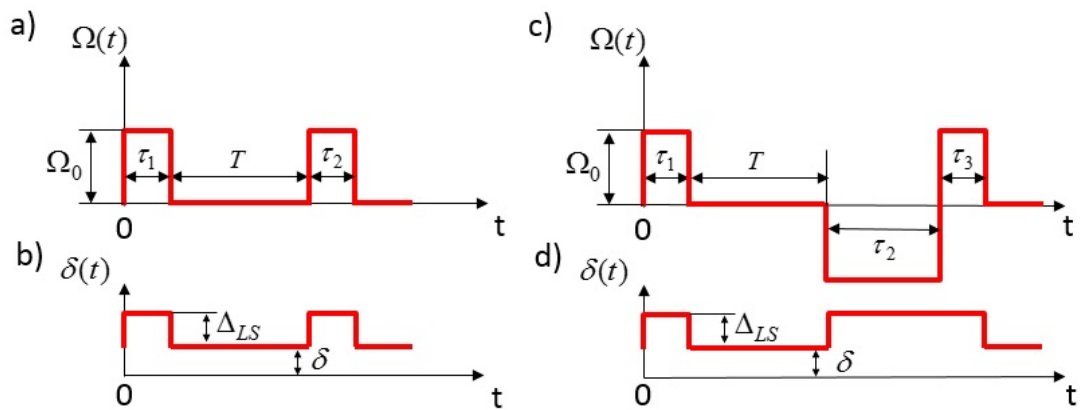

FIG. 2. (Color online) (a) Sequence of laser pulses of the Ramsey method of interrogation at the entrance to the medium. Areas of the pulses are $\Omega_0\tau_1 = \Omega_0\tau_2 = \pi/2$. (c) Sequence of laser pulses of the hyper-Ramsey method of interrogation at the entrance to the medium. Areas of the pulses are $\Omega_0\tau_1 = \Omega_0\tau_3 = \pi/2$, $\Omega_0\tau_2 = -\pi$. (b,d) Dependencies of the detuning on time.

Ramsey resonance for the coordinate z=0 when using a pulse train from Fig.2(c) is shown in Fig.3(a). The dependence of the position of its central minimum on the light shift is a cubic parabola (Fig.3(b), black curve). Let's analyze changes in the hyper-Ramsey resonance and the dependence $S(\Delta_{LS})$ in optically dense medium or thickness (on coordinate z).

Usually in experiments, the lifetime of an excited state of an atom is much longer than the duration of hyper-Ramsey pulse sequence, thus in this section we neglect the terms with spontaneous decay rate γ in the first three equations of the set (7). But the matrix element of dipole moment in the right side of the last equation in the set (7) is not equal to zero, and the absorption effects take place.

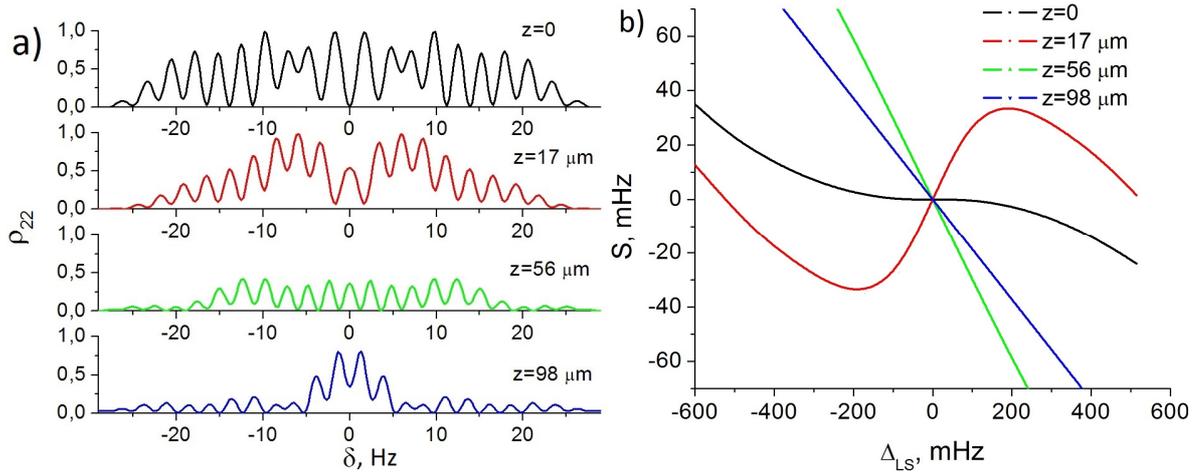

FIG. 3. (Color online) (a) Evolution of the hyper-Ramsey resonance with the longitudinal coordinate (thickness) z along propagation of the radiation. (b) Change in the dependence of the resonance position on the light shift of the atomic transition with the longitudinal coordinate z along propagation of the radiation. Durations of the pulses at the entrance of the medium: $\tau_1 = \tau_3 = \tau$, $\tau_2 = 2\tau$, T=9τ, pulses amplitude $\Omega_0 = (\pi/2)\tau^{-1}$, τ=0.2 s, spontaneous decay rate γ=0. Atomic density $n_a = 10^{11}$ cm$^{-3}$.

In the Fig.3(a) one can see that at the entrance of the medium the central resonance is the minimum. Further, in the cross section z=17 μm it turns into a maximum. Then the process is repeated, and in the cross section z=100 μm there is a minimum again. Thus, population oscillations occur along the z coordinate.

In the Fig.3(b) one can see the dependence $S(\Delta_{LS})$ for the resonance in different cross sections z. This dependence for the maximum in cross section z=17 μm is an N-shaped curve (red curve). In this case, a significant sensitivity to the light shift in the linear portion of the N-shaped curve arises near zero. The least sensitivity to fluctuations of the light shift is achieved in its extremes for the values of the light shift $\Delta_{LS} \approx \pm 200$ mHz. It should be borne in mind that with an increase in the z coordinate, the resonance amplitude of course decreases with the absorption of radiation in the medium.

With a further increase in the z coordinate, the pulse sequence shown in Fig.2(c) is already significantly distorted due to the processes of absorption and re-emission of photons by the atomic medium. This process, in particular, leads to the fact that in the free evolution time the re-emitted field of atoms of the previous layers acts on the atoms in the next layers. The dependence $S(\Delta_{LS})$ becomes linear, as in the classical Ramsey interrogation scheme (Fig.3(b) (green and blue curves)).

We have established that it is possible to modify the pulse areas of the hyper-Ramsey sequence so that a flat part of the dependence $S(\Delta_{LS})$ re-appears at a certain point along the z coordinate in the medium. For example, the pulse sequence shown in the Fig.4 (a) has the areas of the first and the second pulses increased by 10% and the area of the third pulse reduced by 10%.

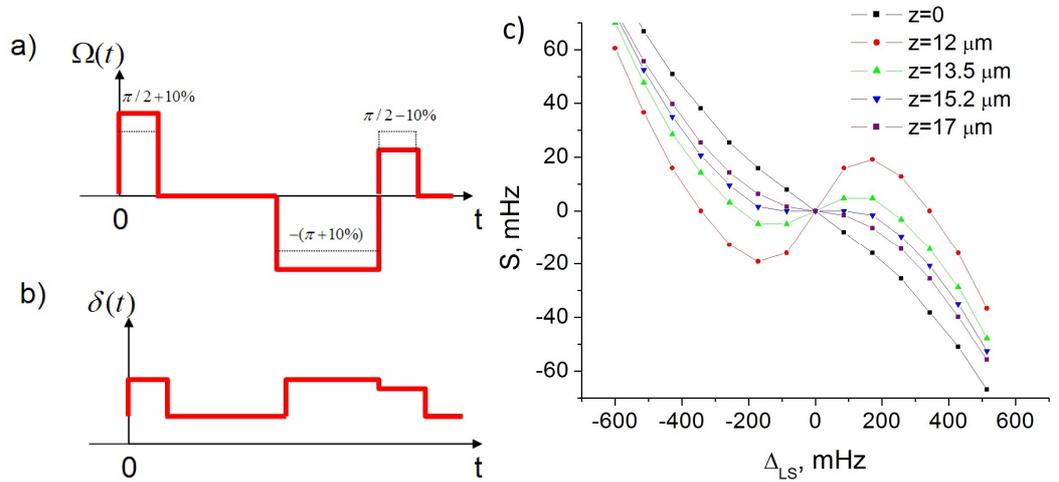

FIG. 4. (Color online) (a) - sequence of hyper-Ramsey pulses when taking into account optical thickness of the atomic medium. (b) Dependence of the detuning on time. (c) Change in the dependence of the resonance position on the light shift of the atomic transition with the longitudinal coordinate z along propagation of the radiation.

Such change in the areas can be achieved by variation of the amplitudes of the pulses. According the formula (8) it leads to the complex dynamics of detuning (Fig.4(b)).

The dependence $S(\Delta_{LS})$ for the pulse sequence from the Fig.4(a) is shown in the Fig.4(c). The curve corresponding to the beginning of the medium (z=0, black curve) is close to a linear dependence. However, it is modified in the depth of the medium in such a way that in the cross section z=15.2 μm (blue curve) the curve has a wide flat plateau in the vicinity of zero. Thus, it is possible to select the required pulse areas for a specific length of the medium.

**3.2 Analyze of the error signal**

It is possible to achieve a lower sensitivity of the atomic resonance shift depending on the light shift of the reference transition when detecting an error signal [29]. In contrast to the fluorescence signal, considered in section 3.1, the error signal is generated when a double interrogation of atoms by a sequence of pulses with a phase-step. Due to the symmetry of the scheme, equal and opposite shifts of the reference resonance appear for each of the two interrogations. During the formation of the error signal, some of them are completely subtracted, which makes the error signal much less sensitive to fluctuations of the laser field. There are several protocols which generate a phase-step error signal [29]. For formal designation, we divide the sequence of laser pulses into four parts. We will denote as $\theta_i$ the area of the $i$-th pulse in degrees, $\varphi_i$ is the initial phase of the $i$-th pulse in radians (Fig.5). $P(\varphi_1, \varphi_3, \varphi_4) = \rho_{22}|_{\text{after the last pulse}}$ is the population of the excited level after interaction of atoms with the sequence of pulses, shown in Fig.5, the initial phases of which have the values $\varphi_1$, $\varphi_3$ and $\varphi_4$. The initial phase $\varphi_2$ in the second region, where the field is missing, is equal to zero. Thus the error signal is

$$\Delta E = P(\varphi_1, \varphi_3, \varphi_4) - P(\varphi_1', \varphi_3', \varphi_4'), \tag{9}$$

where $\varphi_i'$ are the initial phases of the pulses in the secondary interrogation of atoms. The time between interrogations is assumed to be long enough for all atoms to go to the ground state.

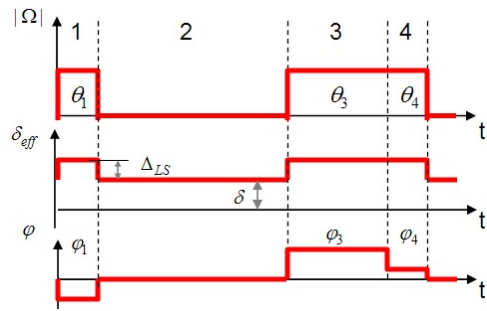

FIG. 5. (Color online) Pulse sequence with arbitrary areas $\theta_i$ and initial phases $\varphi_i$ at the entrance to the medium.

Conventionally, the double pulse sequence is denoted as

$$\theta_{1(\varphi_1,\varphi_1')} - \delta T - \theta_{3(\varphi_3,\varphi_3')}\theta_{4(\varphi_4,\varphi_4')}. \tag{10}$$

Known protocols are shown in the Table 1 [29]. The last three protocols make it possible to completely suppress the sensitivity of the position of the resonance to the light shift of the atomic transition in an optically thin medium.

Table 1. Double interrogation protocols of the composite pulses using a phase-step.

| Protocols | Composite pulses |
| --- | --- |
| Hyper-Ramsey-π | $90_{(\pi/2,-\pi/2)} - \delta T - 180_{(\pi,\pi)}90_{(0,0)}$ |
| Modified hyper-Ramsey | $90_{(\pi/2,0)} - \delta T - 180_{(\pi,\pi)}90_{(0,-\pi/2)}$ |
| Generalized hyper-Ramsey (π/4) | $90_{(0,0)} - \delta T - 180_{(\pi/4,-\pi/4)}90_{(0,0)}$ |
| Generalized hyper-Ramsey (3π/4) | $90_{(0,0)} - \delta T - 180_{(3\pi/4,-3\pi/4)}90_{(0,0)}$ |

The presence of absorption in the medium leads to a significant distortion of the pulse shape. Moreover, this distortion occurs in different ways for the first pulse train with the phases $\varphi_i$ and the second pulse train with the phases $\varphi_i'$, which leads to substantially asymmetrical shifts, whish do not compensate each other when forming the error signal. Thus, protocols that completely suppress the light shift in an optically thin medium require modification in an optically dense medium. Consider further each of them separately. Here we also neglect the terms with spontaneous decay γ in the first three equations in the set (7) on the assumption that the pulses duration is much less than the lifetime of the excited atomic state.

Hyper-Ramsey scheme (HR-π)

Let's consider the dependence of the error signal on the detuning of the laser field for the hyper-Ramsey sequence of pulses $90_{(\pi/2,-\pi/2)} - \delta T - 180_{(\pi,\pi)}90_{(0,0)}$. The error signal is

$$\Delta E_{HR} = P(\pi/2,\pi,0) - P(-\pi/2,\pi,0). \tag{11}$$

In the Fig.6(a) it is shown several graphs of the error signal in various sections along the atomic cloud. When z=0, the signal in the zero offset region is a straight line with a positive slope. (Fig.6(a), black curve). As z increases, the sign of the slope of the straight line changes (Fig.6(a), red and green curves), that is similar to extremum flipping when detecting a fluorescence signal.

In the Fig.6(b) one can see the dependence of the position of the zero point of the error signal on the light shift of the atomic transition. At the entrance to the medium there is a horizontal plateau (black curve). With the z coordinate is increasing, a plateau rotation is observed, similar to the fluorescence signal.

Modified hyper-Ramsey scheme (MHR)

In the work [29] some modifications of the hyper-Ramsey interrogation scheme were proposed, allowing to completely suppress the sensitivity to light shift of the position of the zero point of the error signal in an optically thin medium. Such schemes include a modified hyper-Ramsey scheme, having a formula $90_{(\pi/2,0)} - \delta T - 180_{(\pi,\pi)} 90_{(0,-\pi/2)}$. The error signal for this scheme is

$$\Delta E_{MHR} = P(\pi/2, \pi, 0) - P(0, \pi, -\pi/2). \qquad (12)$$

In the Fig.6(c,d) it is shown the change in the error signal and sensitivity to the light shift of the atomic transition as it moves along the z coordinate for MHR scheme. At the entrance to the medium the sensitivity to light shift is completely absent (Fig.6(d), black curve). However, as the z coordinate increases, the horizontal line rotates, and the effect is stronger than for the HR scheme.

Generalized hyper-Ramsey scheme (GHR)

The error signal for GHR scheme of double interrogation of atoms with arbitrary phase is

$$\Delta E_{GHR(\varphi_3)} = P(0, \varphi_3, 0) - P(0, -\varphi_3, 0). \qquad (13)$$

It is possible to get different versions of this interrogation scheme by setting different values of the phase $\varphi_3$. In the Fig.6(e-h) it is shown the error signals and dependencies of the position of zero point on the light shift for the two cases: GHR($\pi$/4) and GHR(3$\pi$/4). One can see that the horizontal straight lines in the Fig.6(f,h) for z=0 begin to rotate in opposite directions with increasing the z coordinate. The magnitude of the rotation angle is comparable to this value for the MHR scheme.

The opposite directions of plateau rotation for these two schemes has a certain symmetry and makes it possible to build an error signal in whish these effects cancel each other out:

$$\Delta E_{GHR(\pi/4, 3\pi/4)} = \frac{1}{2} (\Delta E_{GHR(\pi/4)} - \Delta E_{GHR(3\pi/4)}). \qquad (14)$$

For this scheme GHR(π/4, 3π/4) it is necessary to interrogate the atoms by a sequence of pulses four times. The signal and the dependence of the zero point on the light shift for this hybrid scheme are shown in the Fig.6(i,j). One can see that the protocol is significantly more resistant to shifts in an optically dense medium. Moreover, in contrast to the method described in section 3.1 allowing to achieve suppression of sensitivity to light shift in a certain point of the medium, this method allows us to suppress the sensitivity to light shifts along the whole medium.

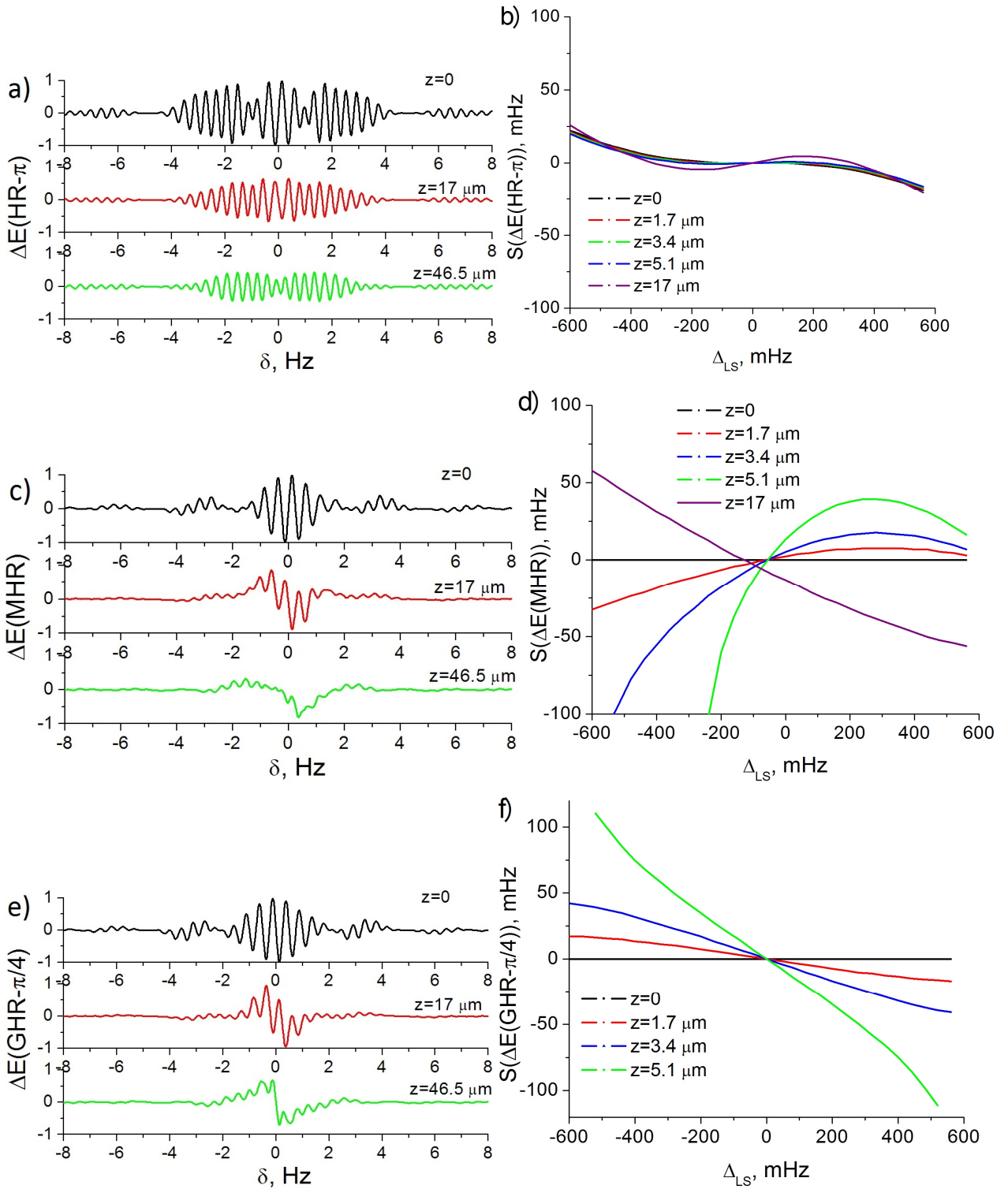

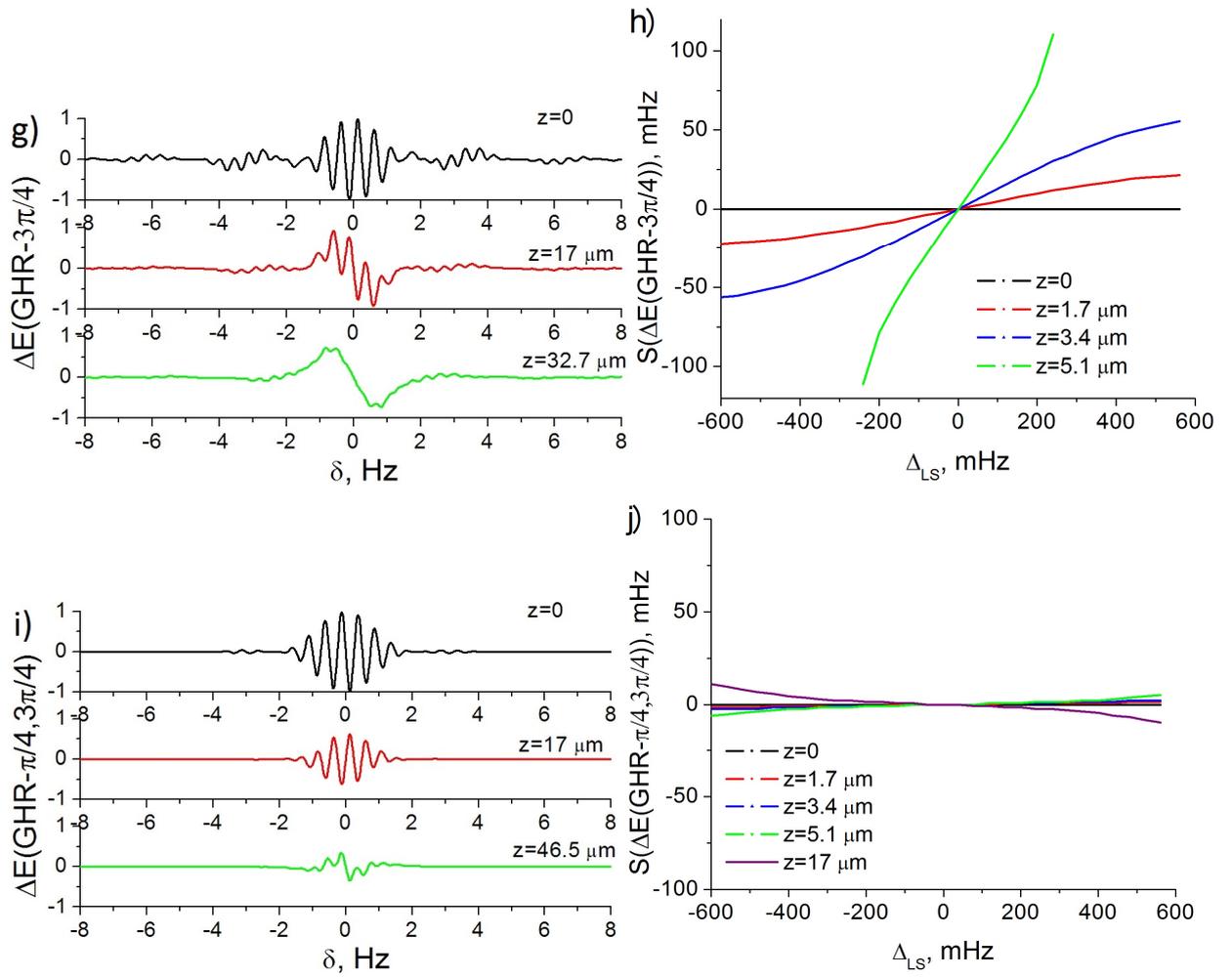

FIG.6 (Color online) (a, c, e, g, i) Error signal and its change with increasing of the longitudinal coordinate z for different schemes. (b, d, f, h, j) Change in the dependence of the position of zero point of the error signal on the light shift of the atomic transition with the longitudinal coordinate z.

Table 2. Angles of plateau rotation of different protocols for the longitudinal coordinate z=5.1 μm.

| Protocols | Angle of plateau rotation for z=5.1 μm |
|---|---|
| HR-π | $0.36^0$ |
| MHR | $14.35^0$ |
| GHR (π/4) | $-9.81^0$ |
| GHR (3π/4) | $19.79^0$ |
| GHR (π/4, 3π/4) | $0.30^0$ |

In the Table 2 angles of rotation of the horizontal part of the dependence of zero point of the error signal on the light shift of the atomic transition are given for the value of the longitudinal coordinate $z=5.1$ μm for all considered protocols. The smallest rotation has the hybrid GHR($\pi/4$, $3\pi/4$) scheme, and comparable result is for HR-$\pi$ scheme. The advantages of GHR($\pi/4$, $3\pi/4$) scheme include the fact that the width of the flat area is wider than that of HR-$\pi$ scheme. However GHR($\pi/4$, $3\pi/4$) is a four-pass scheme, which lengthens the interrogation time of atoms compared to a single-pass HR-$\pi$ scheme.

The rest of considered interrogation schemes (MHR GHR($\pi/4$), GHR($3\pi/4$)) have the much larger (two orders of magnitude) angles of rotation of horizontal section of the dependence of zero point of the error signal on the light shift. It makes these schemes unstable in the case of an optically dense medium, despite the fact that in an optically thin medium they made it possible to completely suppress the dependence on the light shift.

### 3.3 Effect of spontaneous decay of atomic excitation

In the previous calculations we neglected the terms with spontaneous decay rate $\gamma$ in the first three equations in the set (7). This is true in the assumption of short pulses compared with the lifetime of the atomic excitation. But a certain influence on the sensitivity to the light shift still exists. This influence consists in plateau rotation to the same direction as due to optical density of the medium.

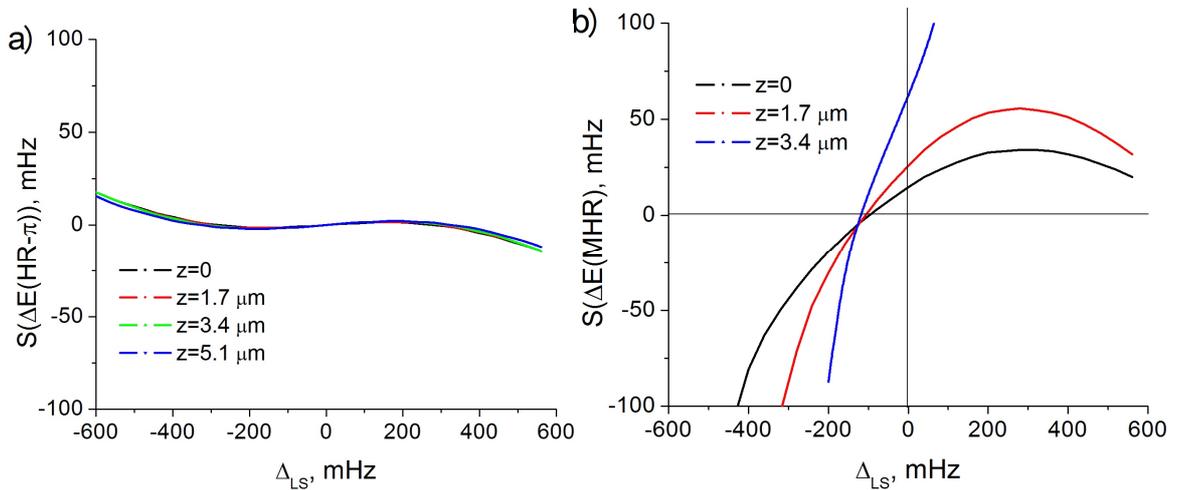

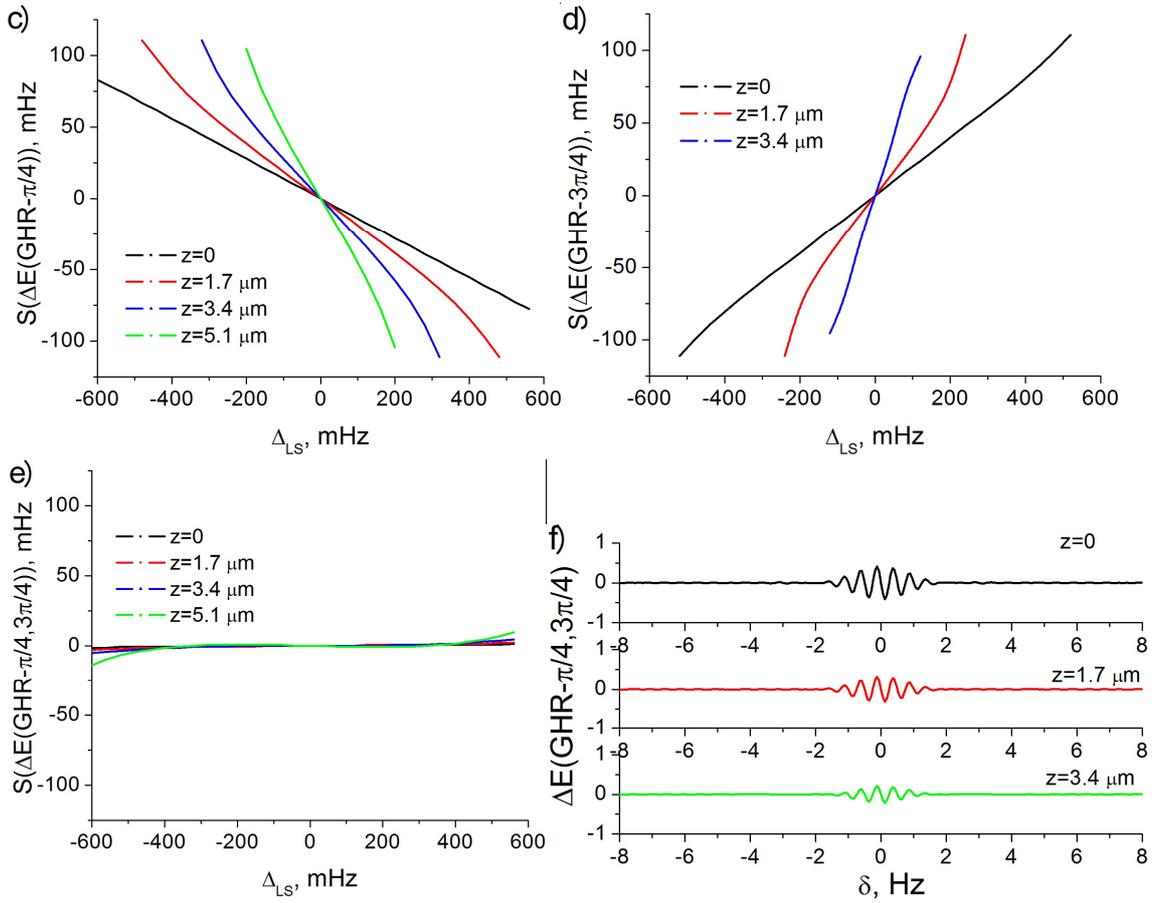

FIG.7 (Color online) (a-e) Change in the dependence of the position of zero point of the error signal on the light shift of the atomic transition with the longitudinal coordinate z. (f) Error signal and its change with increasing of the longitudinal coordinate z for hybrid scheme GHR($\pi/4$, $3\pi/4$). Here spontaneous decay rate of the excited atomic state $\gamma=2\pi*100$ mHz is taken into account.

This is because the effects of absorption of the radiation in optically dense medium and spontaneous decay of atomic excitation lead to energy dissipation from the system. In the first case, energy is lost to the excitation of the atoms of the previous layers, and in another case to the filling of vacuum modes of radiation. Therefore, the graphs in the Fig.7 (a-e) are similar to the graphs in the Fig.6 (b, d, f, h, j) respectively.

Thus, the hybrid scheme GHR($\pi/4$, $3\pi/4$) also works in this case of spontaneous decay (the error signal is formed by the formula (14)). The dependence of the position of zero point of the error signal on the light shift of the atomic transition remains horizontal (Fig.7(e)). But the amplitude of the error signal is less compared to the signal in the absence of spontaneous decay (comparison of the black curves in the Fig.6(i) and Fig.7(e)). In addition, in the presence of spontaneous decay, the amplitude decreases with the longitudinal coordinate z much faster.

## 4. Conclusion

In this work the theory of the hyper-Ramsey interrogation scheme has been developed for "two-level" atom under conditions of finite optical thickness and the presence of the collective effects in dilute medium. The cold atomic ensemble is considered as a medium. The mathematical model is the set of Maxwell-Bloch equations which consists of dynamic equations for the atomic density matrix and equation of propagation of the electromagnetic field. In the framework of the developed model it is found that the shape of the hyper-Ramsey resonance changes significantly due to absorption in the atomic medium.

The different protocols of hyper-Ramsey composite interrogation pulse trains were analyzed.

For the simplest case of fluorescence signal of hyper-Ramsey-$\pi$ protocol it was shown that the shape of the central fringe changed from dip to peak and vice versa due to an additional pulse area accumulation during the entire interrogation sequence. The analysis of the light shift showed that its dependence on the detuning is cubic only for a thin medium ($z=0$). As it passes through the medium, this cubic dependence changes and becomes linear. It is found that the modification of the pulse areas in the hyper-Ramsey sequence, namely, increasing the areas of the first and second pulses by 10% and reducing the third pulse by 10%, makes it possible to obtain a wide plateau on the dependence of the position of the resonance on light shift at a certain point inside the medium.

Next we analyzed the error signal for hyper-Ramsey, modified and generalized hyper-Ramsey protocols. Despite the fact that these methods make it possible to fully compensate for the sensitivity to the light shift, even a small absorption in the medium leads to a strong dependence of the resonance on the light shift. This is expressed as the rotation of the flat area on the graph of the dependence of the shift of the resonance on the light shift of the atomic transition. However, the use of a hybrid generalized hyper-Ramsey scheme (atoms are interrogated 4 times) allows us to compensate this rotation. Thus, a method of suppression of sensitivity of an atomic resonance to the light shift in any point in a dense medium was found.

Finally, the effect of spontaneous decay of excited atomic state was analyzed. It was shown that spontaneous decay and absorption of the radiation in optically dense medium have the similar effect (due to energy dissipation) and lead to rotation of plateau in the curves of sensitivity to the light shift in the same direction.

**Acknowledgments**

This work was supported by Russian Foundation for Basic Research (project № 18-32-20022 mol_a_ved).

We would like to acknowledge Denis V. Brazhnikov for discussions and suggestions.**References**

[1] I.I. Rabi // Phys. Rev. **51**, 652 (1937)

[2] I.I. Rabi, J.R. Zacharias, S. Millman and P. Kusch // Phys. Rev. **53**, 318 (1938)

[3] L. Allen, J.H. Eberly "Optical resonance and two-level atoms"

[4] T. Rosenband, et. al. // Science **319**, 1808–12 (2008)

[5] H.S. Margolis // Eur. Phys. J. Spec. Top. **172** 97 (2009)

[6] C.W. Chou, D.B. Hume, J.C.J. Koelemeij, D.J. Wineland and T. Rosenband // Phys. Rev. Lett. **104** 070802 (2010)

[7] J. Ye, H.J. Kimble and H. Katori // Science **27**, 1734 (2008)

[8] A. Derevianko and H. Katori // Rev. Mod. Phys. **83** 331 (3022)

[9] H. Katori // Nat. Photon. **5**, 203 (2011)

[10] E. Tkalya // JETP Lett. **71**, 311 (2000)

[11] G.A. Kazakov, A.N. Litvinov, V.I. Romanenko, L.P. Yatsenko, A.V. Romanenko, M. Schreitl, G. Winkler and T. Schumm // New Journal of Physics **14**, 083019 (2012)

[12] E. Peik, Chr. Tamm // Europhysics Letters **61**, No. 2 (2003)

[13] M. S. Safronova, D. Budker, D. DeMille, Derek F. Jackson Kimball, A. Derevianko, and Charles W. Clark // Rev. Mod. Phys. **90**, 025008 (2018)

[14] V. I. Yudin and A. V. Taichenachev // arXiv:1706.07718v3 (2017)

[15] A. Derevianko and M. Pospelov // Nature Phys. **10**, 933 (2014)

[16] A.D. Ludlow, M.M. Boyd, J. Ye, E. Peik and P.O. Schmidt // Rev. Mod. Phys. **87**, 637 (2015)

[17] S. M. Brewer, J.-S. Chen, A. M. Hankin, E. R. Clements, C. W. Chou, D. J. Wineland, D. B. Hume, and D. R. Leibrandt // Phys. Rev. Lett. **123**, 033201 (2019)

[18] N. F. Ramsey // Phys. Rev. **78**, 695 (1950)

[19] L. Essen and J.V.L. Parry // Nature **176**, 280 (1955)

[20] J. Vanier and C. Audoin "The Quantum Physics of Atomic Frequency Standards", ed. A Hilger (Bristol: IOP) (1989)

[21] C.J. Bordé // Physics Letters A **140**, Iss.1-2, pp 10-12 (1989)

[22] C.J. Bordé // Métrologia **39**, 435, (2002)

[23] V. I. Yudin, A. V. Taichenachev, C. W. Oates, Z. W. Barber, N. D. Lemke, A. D. Ludlow, U. Sterr, Ch. Lisdat, and F. Riehle // Phys. Rev. A **82**, 011804 (2010)